\documentclass[12pt]{article}
\usepackage{times}
\usepackage{epsfig}
\usepackage{color}
\usepackage{sublabel}
\usepackage[numbers,sort&compress,round]{natbib}

\makeatletter
  \def\@eqnnum{{\normalfont \normalcolor [\theequation]}}
\makeatother

\title{\vspace*{-20mm}\sffamily\textbf{The Evolution of Extortion \\
in Iterated Prisoner's Dilemma Games}}
\author{Christian Hilbe$^1$, Martin A. Nowak $^2$,  \& Karl Sigmund $^{3,4}$\\
\vspace{-2mm}\normalsize$^1$ Max Planck Institut for Evolutionary Biology, Ploen\\
\vspace{-2mm}\normalsize $^2$ Program for Evolutionary Dynamics, Harvard University,\\
\vspace{-2mm}\normalsize One Brattle Square, Cambridge, MA 02138, USA\\
\vspace{-2mm}\normalsize$^3$ Faculty of Mathematics, University of Vienna, A-1090 Vienna, Austria\\
\vspace{-2mm}\normalsize$^4$ International Institute for Applied Systems Analysis, A-2361 Laxenburg, Austria}
\date{\normalsize\today}

\begin{document}
\maketitle


\subsubsection*{Corresponding author:}
{\small Karl Sigmund\\
Faculty of Mathematics, University of Vienna, A-1090 Vienna, Austria\\
e-mail: karl.sigmund@univie.ac.at, phone: +43 (0)1 4277 506 12, fax: +43 (0)1 4277 9 506}

\newpage
{\bf 
Iterated games are a fundamental component of economic and evolutionary game theory.
They describe situations where two players interact repeatedly and have the possibility to
use conditional strategies that depend on the outcome of previous interactions. In the context
of evolution of cooperation, repeated games represent the mechanism of reciprocation. 
Recently a new class of strategies has been proposed, so called `zero determinant strategies'.
These strategies enforce a fixed linear relationship between one's own payoff and that of the other 
player. A subset of those strategies are `extortioners' which ensure that any increase in the own payoff exceeds
that of the other player by a fixed percentage. Here we analyze the evolutionary performance
of this new class of strategies.  We show that in reasonably large populations they can act as catalysts 
for the evolution of cooperation, similar to tit-for-tat, but they are not the stable outcome of natural selection. 
In very small populations, however, relative payoff differences between two players in a contest matter, 
and extortioners hold their ground. Extortion strategies do particularly well in co-evolutionary arms races between two distinct 
populations: significantly, they benefit the population which evolves at the slower rate - an instance of the so-called  Red King effect.
This may affect the evolution of interactions between host species and their endosymbionts.} 

\section*{Introduction}

The Iterated Prisoner's Dilemma ($IPD$) has a long history of serving as a model for the cultural and biological evolution of cooperation \cite{rapoport65,trivers71,aumann81,axelrod84,fudenberg86,fudenberg90,nowak93,kendall07,trivers06}. A new class of so called {\it zero-determinant (ZD)} strategies has recently attracted considerable attention \cite{press12,stewart12, adami12}. Such strategies allow players to unilaterally enforce a linear relation between the own and the co-player's payoff. A subset consists of the so-called {\it equalizer} strategies: these assign to the co-player's score a predetermined value, independent of the co-player's strategy, see also \cite{boerlijst97}. Another subset consists of the {\it extortion} strategies: they guarantee that the own surplus exceeds the co-player's surplus by a fixed percentage. Press and Dyson \cite{press12}  have explored the power of $ZD$-strategies to manipulate any 'evolutionary' opponent, i.e., any co-player able to learn, and to adapt.

In their commentary to Press and Dyson, Stewart and Plotkin \cite{stewart12} ask: 'What does the existence of $ZD$ strategies mean for evolutionary game theory: can such strategies naturally arise by mutation, invade, and remain dominant in evolving populations?' From the outset, it may seem that the opportunities for extortion strategies are limited. If a strategy is successful, it will spread, and therefore be more likely to be matched against its like: but any two extortioners hold each other down to surplus zero. However, if the two players engaged in an $IPD$ belong to distinct populations, the evolutionary prospects of extortion improve significantly.  
 
In the following, we investigate the impact of $ZD$-strategies on evolutionary game theory. We show that in well-mixed populations, $ZD$-strategies can play an important role, but only as catalyzers, not as long-term outcome. However, if the $IPD$ is played between members of two {\it separate} populations evolving on different time-scales, extortion strategies can get the upper hand in whichever population evolves more slowly, and enable it to enslave the other population, an interesting example of the so-called Red-King effect \cite{bergstrom03}.    

The Prisoner's Dilemma ($PD$) game is a game between two players $I$ and $II$ having two strategies each, which we denote by $C$ ('to cooperate') and $D$ ('to defect'). It is assumed that the payoff for two cooperating players, $R$, is larger than the payoff for two defecting players, $P$. If one player cooperates and the other defects, the defector's payoff $T$ is larger than $R$, and the cooperator's payoff $S$ smaller than $P$. Thus the game is defined by $ T>R>P>S $. An important special case is the so-called donation game, where each player can 'cooperate' (play $C$) by providing a benefit $b$ to the other player at own cost $c$, with $0<c<b$. Then $T=b$, $R=b-c$, $P=0$ and $S=-c$. 

In the Iterated Prisoner's Dilemma game ($IPD$), the two players are required to play an infinite number of rounds, and their payoffs $P_I$ resp. $P_{II}$ are given by the limit in the mean of the payoffs per round. An important class of strategies consists of so-called {\it memory-one} strategies. They are given by the conditional probabilities $p_R, p_S, p_T $ and $p_P$ to play $C$ after experiencing outcome $R,S,T$ resp. $P$ in the previous round. (In addition, such a strategy has to specify the move in the first round, but this plays no role in the long run, see e.g. \cite{sigmund10}). 
An important class of memory-one strategies consists of {\it reactive} strategies, which only depend on the co-player's move in the previous round (not the own). Then $p_R=p_T=:p$ and $p_P=p_S=:q $, so that a reactive strategy corresponds to a point $(p,q)$ in the unit square \cite{nowak90}.

We will first define and characterize zero-determinant strategies, equalizers and extortioners. We then investigate, in the context of evolutionary game theory, the contest between extortioners and four of the most important memory-one strategies. We will show that extortion cannot be an outcome of evolution, but can catalyze the emergence of cooperation. The same result will then be obtained if we consider {\it all} memory-one strategies: in particular, $ZD$-strategies can only get a foothold if the population is very small. If the $IPD$ is played between members of two distinct populations, $ZD$-strategies can emerge in the population which evolves more slowly. In particular, extortion strategies can allow host species to enslave their endosymbionts.

\section*{Methods and Results}
 
{\bf Definitions.} Press and Dyson \cite{press12} define the class of 'zero-determinant' strategies $ZD$ as those memory-one strategies $(p_R, p_T, p_S, p_P)$ satisfying, for some reals $\alpha, \beta , \gamma $, the equations
\sublabon{equation}
\begin{eqnarray}
p_R-1 &= &\alpha R + \beta R + \gamma \\ 
p_S-1 &= & \alpha S + \beta T + \gamma\\
p_T &= & \alpha T + \beta S + \gamma\\
p_P &= &\alpha P + \beta P + \gamma .
 \end{eqnarray}
 \sublaboff{equation}
(We note that $1-p_R$ and $1-p_S$ are the probabilities to switch from $C$ to $D$, while $p_T$ and $p_P$ are the probabilities to switch from $D$ to $C$.) Press and Dyson showed that if player $I$ uses such a $ZD$ strategy, then 
\begin{equation}
\alpha P_I+\beta P_{II}+\gamma =0, 
\end{equation}
no matter which strategy player $II$ is using.
Equalizer strategies are those $ZD$ strategies for which $\alpha =0 \neq \beta $: then 
\begin{equation}
P_{II}= - \gamma / \beta .
\end{equation}
Thus player $I$ can assign to the co-player any payoff between $P$ and $R$. (Indeed,  since the $p_i$ have to be between $0$ and $1$, it follows that $\beta <0$ and $P \leq P_{II} \leq R$). The so-called $\chi$-extortion strategies are those $ZD$-strategies for which $\gamma =-(\alpha +\beta)P$, with $\chi =-\beta /\alpha > 1$.   Then
\[ P_I-P = \chi (P_{II}-P) . \]  
 In this case, player $I$ can guarantee that the own 'surplus' (over the maximin value $P$) is the $\chi $-fold of the co-player's surplus.
 
 Press and Dyson speak of zero-determinant strategies because they use for their proof of [2] an ingenious method based on determinants. In Appendix A, we present a more elementary proof, following \cite{boerlijst97}. Within the four-dimensional unit cube of all memory-one strategies $(p_R,p_S,p_T,p_P)$, the $ZD$ strategies form a three-dimensional subset  $\mathcal{ZD}$ containing the two-dimensional subsets $\mathcal{EQ}$ and $\mathcal{EX}$ of equalizers resp. extortioners (see Appendix B). In Fig.1 we sketch these sets for the reactive strategies.

{\bf Extortion within one population.} In order to investigate the role of extortion in the context of evolutionary games, we concentrate on the donation game, but stress that the main results hold more generally.   
We first consider how a $\chi$-extortion strategy $E_{\chi}$ fares against some of the most important memory-one strategies, namely  $TFT$ $=(1,0,1,0)$, $AllD$ $=(0,0,0,0)$, $All C$ $=(1,1,1,1)$ and the Win-Stay-Lose-Shift strategy $WSLS$ which is encoded by $(1,0,0,1)$, and hence cooperates if and only if the co-player's move, in the previous round, was the same than the own move, see \cite{nowak93}. For the donation game, the payoff for a player using strategy $i$ against a player with strategy $j$ is given by the $(i,j)$-th element of the following matrix:

\begin{equation} \label{five}
\begin{array}{c|ccccc}
	&TFT	&WSLS	&E_\chi	&All~C	&All~D\\
	\hline
TFT	&(b-c)/2	&(b-c)/2	&0	&b-c	&0\\
WSLS	&(b-c)/2	&b-c &\frac{b^2-c^2}{b(1+2\chi)+c(2+\chi)}	& (2b-c)/2	&-c/2\\
E_\chi	&0	&\frac{(b^2-c^2)\chi}{b(1+2\chi)+c(2+\chi)} &0 &\frac{(b^2-c^2)\chi}{b\chi+c}	&0\\
All~C	&b-c	& (b-2c)/2	&\frac{b^2-c^2}{b\chi+c}	&b-c	&-c\\
All~D	&0	&b/2	&0	&b	&0\\
\end{array}
\end{equation}

Let us start with the pairwise comparisons. $E_{\chi}$ is neutral with respect to $All D$. It is weakly dominated by $TFT$. (A $TFT$-player does not better than an extortioner against extortioners, but interactions with other $TFT$-players are giving an advantage to $TFT$.) $AllC$ players can invade extortioners, and vice versa: these two strategies can stably coexist in proportions $c(\chi - 1):(b+c)$. Finally, $WSLS$ dominates extortioners. We note that the mixed equilibrium of extortioners and unconditional cooperators can be invaded by each of the other three strategies. The same holds for the mixed equilibria of extortioners and unconditional defectors, if the frequency of extortioners is sufficiently high. In particular, $TFT$ can always invade such a mixed equilibrium, but can, in turn, be invaded by $WSLS$ or $AllC$. No Nash equilibrium involves $E_{\chi}$. If $b<2c$, there are two Nash equilibria: a mixture of $TFT$, $AllC$ and $AllD$ and a mixture of $WSLS$, $AllC$ and $AllD$. If $b>2c$, there exist four Nash equilibria. In particular, $WSLS$ is then a strict Nash equilibrium.

The replicator dynamics (see, e.g.,  \cite{helbing92, nowak06, sigmund10}) displays for the payoff matrix \label{five} continuous families of fixed points and periodic orbits, and hence is far from being robust. The same applies to most other deterministic game dynamics. It seems more reliable to consider a stochastic process which describes a finite, well-mixed population consisting of $M$ players, and evolving via copying of successful strategies and exploration, i.e., by a selection-mutation process (see, e.g., \cite{nowak06, nowak04, traulsen06}).  Selection is modeled as an imitation process; in each time step, a randomly chosen individual $A$ adopts the strategy of a role model $B$ with a probability which increases with the model's  payoff.
Additionally, mutations occur with a small probability $\mu>0$ (corresponding to the random adoption of another strategy), thereby ensuring that all population states can be reached by the evolutionary dynamics.

Any such stochastic process yields a steady state distribution of strategies. We find that while extortioners are never the most abundant strategy, they can play the role of a catalyzer. Indeed, if only $All D$ and $WSLS$ are available, a population may be trapped in a non-cooperative state for a considerable time, leading to a mutation-selection equilibrium that clearly favors defectors (see Fig. 2A). In such a case, extortioners (Fig. 2B) and $TFT$ (Fig. 2C) offer an escape: both strategies can subvert an $AllD$ population through neutral drift. Once defectors are rare, $WSLS$ outperforms $TFT$, and it also prevails against extortioners if the population is sufficiently large (in a direct competition, $WSLS$ always gets a higher payoff than $E_\chi$ if $M>1+\chi$). Thus, in large populations, extortioners and $TFT$ tip the mutation-selection balance towards $WSLS$. In contrast, expanding the strategy space by adding $All C$ has a negligible impact on the equilibrium (Figs. 2D and 2E), see also \cite{imhof05}.


What happens when players are not restricted to the five specific strategies, but can choose among all possible memory-one strategies? We study this by using the stochastic evolutionary dynamics of \cite{imhof10}, assuming that mutants can pick up any memory-one strategy, and that the mutant reaches fixation, or is eliminated, before the next mutation occurs. Overall, this stochastic process leads to a sequence of monomorphic populations. The evolutionary importance of a given strategy can then be assessed by computing how often the state of the population is in its neighborhood. 
For a subset $A$ of the set of memory-one strategies, we denote the $\delta$-neighborhood of $A$ (with respect to Euclidean distance) by $A_\delta $, and let $\mu (A_\delta)$ denote the fraction of time that the evolving population visits $A_\delta$. 
We say that $A_\delta$ is favored by evolution if the evolutionary process visits $A_\delta$ more often than expected under neutral evolution, 
i.e., if $\mu (A_\delta)$ is larger than the volume of  $A_\delta$. We apply this concept to $A=\mathcal{ZD},\mathcal{EQ},\mathcal{EX}$.


Extensive simulations indicate that neither extortioners, nor equalizers or \linebreak
zero-determinant strategies, are favored by evolution if the population is reasonably large (see Fig. 3A). By contrast, very small population sizes promote the evolution of these behaviors. For extortioners, this result is intuitive: in small populations,  relative payoff advantages determine the fate of a strategy, rather than absolute payoffs \cite{nowak04}; this effect may even result in the evolution of spite, see \cite{rand10, hilbe12}. It is less clear why the other two behaviors are abundant in small populations, and why large population sizes lead to their decrease. However, simulations show that the decline of zero-determinant strategies and equalizers is partly due to the success of $WSLS$-like strategies, which soon dominate the evolutionary dynamics (see Fig. 3B): as the population size increases, individuals prefer strategies that cooperate after mutual cooperation and after mutual defection, and that defect otherwise. As a (possibly surprising) consequence, larger populations also yield higher average payoffs (Fig. 3C). These qualitative results are robust with respect to changes in parameter values, such as benefits and costs, or the strength of selection, indicating that $WSLS$, rather than extortion, is favored by evolution as soon as the population size exceeds a critical level.

{\bf Extortion between two populations.} Let us now consider two species (for instance, hosts and their symbionts), or two classes of a single species, old and young, for example, buyers and sellers, or rulers and subjects, engaged in an $IPD$ game which, of course, is now unlikely to be symmetric. In such situations, extortioners may evolve even in large populations. Indeed, extortioners provide incentives to cooperate: as shown by Press \& Dyson \cite{press12}, $AllC$ is always a best response to an extortion strategy. In a single population of homogeneous players, this is not turned to advantage, as the extortioners' success leads to more interactions with their own kind. If extortioners evolve in one of two separate populations, they will not have to interact with co-players of their own kind. Nevertheless, their success may be short-lived, since they will be tempted to adopt the even more profitable $AllD$-strategy as a reaction to the $AllC$ co-players which they have produced.

To achieve lasting success in a two-population set-up, extortioners need to be stubborn, and cling to their strategy. To elucidate this point, we extend our previous analysis by revisiting a co-evolutionary model of Damore and Gore \cite{damore11}. These authors consider host-symbiont interactions where each host interacts with its own subpopulation of endo-symbionts.  Let us assume that these interactions are given by an $IPD$ game. 
Members of both species reproduce with a probability proportional to their payoffs, replacing a randomly chosen organism of their species. However, the two populations of hosts and symbionts may evolve on different time scales, as measured by their relative evolutionary rate ($RER$). For a relative evolutionary rate of one, hosts and symbionts evolve at a similar pace in the evolutionary arms race, and no population is able to extort the other (Fig. 4A). This changes drastically as soon as we increase the relative evolutionary rate, by allowing symbionts to adapt faster. Fast adaptation results in a short-term increase of the symbionts' payoffs, since they can quickly adjust to their respective host. In the long term, however, hosts learn to adopt extortion strategies (Fig. 4B), thereby forcing their symbionts to cooperate. Thus it pays in the long run, for the host, to be pigheaded and refuse to adapt; for the parameters in Fig. 4B, the resulting equilibrium allocates them on average more than $90 \%$ of the surplus.

\section*{Discussion}

Our main results show that within one population, extortioner strategies can act as catalyzers for cooperation, but prevail only if the population size is small; and that in interactions between two populations, extortion can emerge if the rates of evolution differ. This holds not only for the donation game (and therefore whenever $R+P=T+S$), but in considerably more general contexts. We could assume, for instance, that the players alternate their moves in the donation game \cite{nowak94, frean94}; or that the underlying $PD$ game is asymmetric (the definitions have to be modified in an straightforward way). As noted in \cite{press12}, some results hold also for non-$PD$ games; this deserves further investigation.

Extortion strategies are only a small subset of $ZD$-strategies. We have shown that within large populations, the class of $ZD$ strategies is not favored by selection, in the sense that its neighborhood is not visited dis-proportionally often. This does not preclude, of course, that certain elements of this class are favored by selection. Thus Generous TFT $(1, 1-c/b,1,1-c/b)$ does well. So do other, less known strategies. In particular, Stewart and Plotkin highlighted a class of strategies defined, instead of Eq. [3], by $P_I-R=\chi (P_{II}-R)$ (with $\chi >1)$). A player using this strategy does not claim a larger portion of the 'surplus', but a larger share of the 'loss' (relative to the outcome $R$ of full cooperation). Remarkably, these 'compliant' strategies do as well as $WSLS$.

In a preprint by Adami and Hintze \cite{adami12}, the evolutionary stability of several $ZD$ strategies was tested by replicator dynamics and agent-based simulations, which independently confirms the result that these strategies do not prevail in large populations. They used payoff values $R=3$, $S=0$, $T=5$ and $P=1$, i.e., a Prisoner's Dilemma game which cannot  be reduced to a donation game. Adami and Hintze also discuss the evolutionary success of 'tag-based' strategies, which use extortion only against those opponents who do not share their tag. These
strategies are not memory-one strategies, but use them in specific contexts.

In \cite{press12},  Press and Dyson analyzed $ZD$ strategies in the context of classical game theory, with two players locked in contest: extortion strategies play an important role,
as do the more orthodox trigger strategies, see \cite{aumann81, fudenberg90}. In the context of evolutionary game theory, whole populations are engaged in the game. Not surprisingly, for very small population size extortion strategies still offer good prospects. In larger populations (with our parameter values, for $M>10$), this is no longer the case. However, evolutionary game theory can reflect features of classical game theory if the two interacting players game belong to two separate evolving populations. 

Therefore, extortion may evolve in endosymbiotic relationships due to the so-called Red-King effect \cite{bergstrom03, frean04, doebeli98}: the species that evolves at a slower rate gains a disproportionate share of the benefits. This requires two conditions to be met: individuals need to come from different populations, and these populations have to evolve on different time scales. If these conditions are fulfilled, extortioner hosts can manipulate their symbionts' evolutionary landscape in such a way that the host's and the symbionts' payoffs are perfectly correlated. This ensures that only those symbiont mutants can succeed that are beneficial for the host. In this sense, such hosts apply an evolutionary kind of mechanism design; they create an environment that makes the symbionts' cooperation profitable for the symbionts, but even more profitable for themselves.\\

\noindent {\bf Appendix A: Proof of Eq. [2]}  Let us denote by $P_I(n)$ and $P_{II}(n)$ the player's payoffs in round $n$, by $s_i(n)$ the probability that $I$ experiences outcome $i \in \{R,S,T,P \}$ in that round and by $q_i(n)$ the conditional probability, given that outcome, that  $II$ plays $C$ in round $n+1$. By conditioning on round $n$, we see that $s_R(n+1)$ is given by 
\[ s_R(n)q_R(n)p_R+s_S(n)q_S(n)p_S+s_T(n)q_T(n)p_T+s_P(n)q_P(n)p_P , \]
and $s_S(n+1)$ by
\[ s_R(n) (1-q_R(n))p_R+s_S(n)(1-q_S(n))p_S+s_T(n)(1-q_T(n))p_T+s_P(n)(1-q_P(n))p_P .\]
Using (1), and setting ${\bf g}_I:=(R,S,T,P)$, ${\bf g}_{II}:=(R,T,S,P)$,  we see that $w(n):= s_R(n+1)+s_S(n+1)-(s_R(n)+s_S(n))$ is given by
\[ \alpha {\bf s}(n) \cdot {\bf g}_I+\beta {\bf s}(n) \cdot {\bf g}_{II}+\gamma {\bf s}(n) \cdot {\bf 1} \]
which is just $\alpha P_I(n)+\beta P_{II}(n)+\gamma $. Summing $w(n)$ over $n=0,1,..., N-1$ and dividing by $N$, we see that 
\[ \frac {s_R(N)+s_S(N)-s_R(0)-s_S(0)}{N} \to \alpha P_I+\beta P_{II}+\gamma \] 
and hence Eq. [2] holds, independently of the strategy of player $II$. The same proof works for any $2 \times 2$ game (even if it is asymmetric: one just has to replace ${\bf g}_{II}$ with the corresponding payoff vector). In many cases, however, there will be no solutions to Eq. [1] which are feasible (i.e., probabilities between $0$ and $1$).\\

\noindent {\bf Appendix B: the sets $\mathcal{ZD}$, $\mathcal{EQ}$ and $\mathcal{EX}$:}   Elementary algebra shows that within the four-dimensional unit cube of all memory-one strategies $(p_R,p_S,p_T,p_P)$, the $ZD$-strategies are characterized by
\[ (1-p_R) (S+T-2R)+(1-p_S)(R-P)+p_T(P-R)+p_P(S+T-2P)=0 ,\]
(a three dimensional subset of the cube). Equalizers are characterized, in addition, by
\[ (R-P)(p_S-p_T-1)=(T-S)(p_R-p_P-1), \]
(they form a two-dimensional set) and $\chi $-extortion strategies by $p_P=0$ and
\[ p_T[S+\chi (T-P)]=(1-p_S)[T+\chi (P-S)],  \]
(for each $\chi $ a one-dimensional set). In the special case of the donation game, these equations reduce to
\[ p_R+p_P=p_S+p_T , \]
 \[ (b-c)(p_S-p_T-1)=(b+c)(p_R-p_P-1),\]
\[ p_T(c+\chi b)=(1-p_S)(b+\chi c),\] 
respectively. The set $\mathcal{EQ}$ of equalizers is spanned by $(1,1,0,0)$, $(c/b,0,c/b,0)$, $(\frac {2c}{b+c},0,1,\frac {b-c}{b+c})$ and $(1,1-c/b,1,1-c/b)$, the set $\mathcal{EX}$ of extortion strategies by $(1,1,0,0)$, $(c/b,0,c/b,0)$ and $(1,0,1,0)$. All reactive strategies are $ZD$-strategies, the reactive equalizers are those satisfying $p-q=c/b$, and the reactive $\chi $-extortioners those with $q=0$ and $p=(b+\chi c)/(c+\chi b)$ (see Fig.1). \\

\noindent {\bf Acknowledgement:} KS acknowledges support by Grant \#RFP-12-21 from Foundational Questions in Evolutionary Biology Fund.

\newpage

\begin{figure}[h!]
\centering
\includegraphics[height=5.5cm]{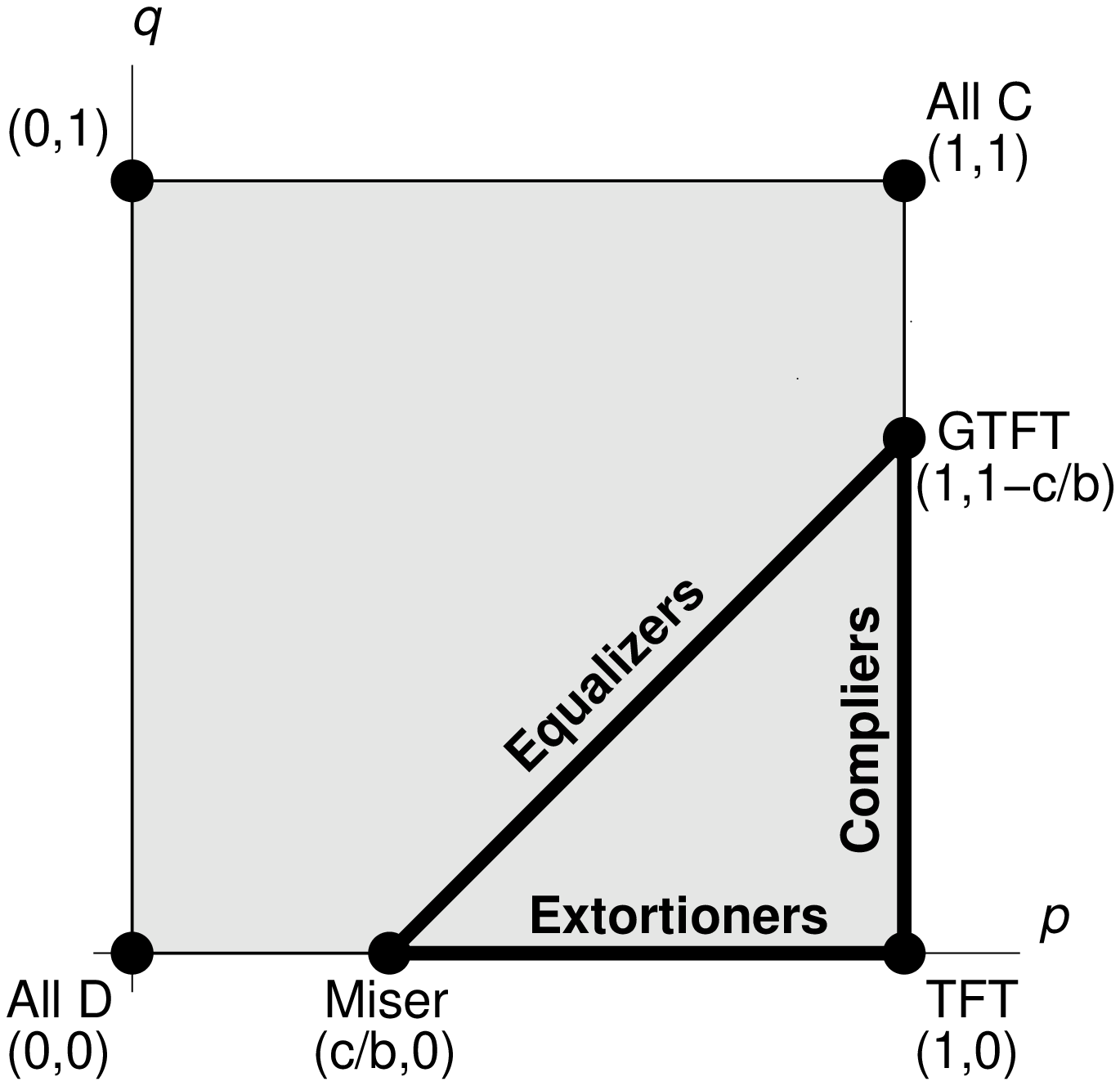}
\caption{Reactive strategies ($p_R=p_T=p$, $p_S=p_P=q$) for the donation game. All reactive strategies (the square $0 \leq p,q \leq 1$)  are $ZD$ strategies. The equalizer strategies are those on the segment between 'generous $TFT$' ($p=1$, $q=1-c/b$, see \cite{nowak90}) and 'Miser' ($p=c/b$, $q=0$,   see \cite{frean94}), the extortion strategies those between 'Miser' and $TFT$ ($p=1$, $q=0$), and the 'compliant' strategies (see \cite{stewart12} and Discussion) those between 'generous $TFT$' and $TFT$.}
\end{figure}
~\\[1cm]

\begin{figure}[h!]
\centering
\includegraphics[height=5.3cm]{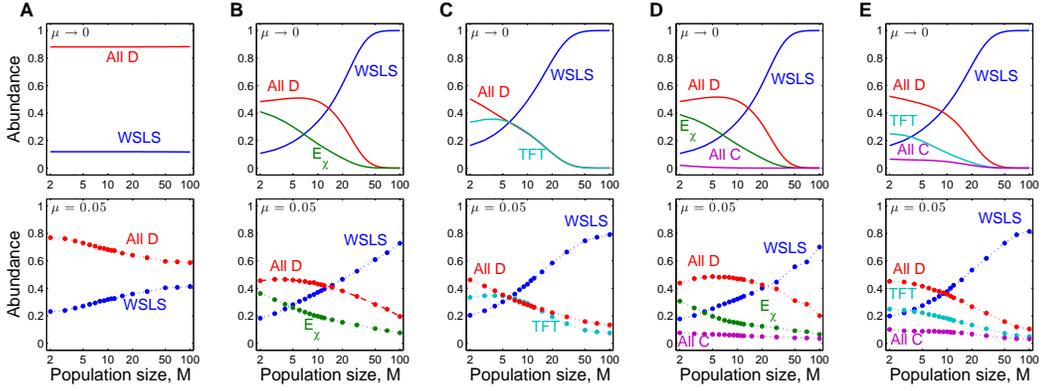}
\caption{Evolutionary competition between some important strategies in the $IPD$. For various population sizes $M$, the graphs show the frequency of each strategy in the mutation-selection equilibrium. We consider two mutation regimes, the limit of rare mutations $\mu\rightarrow 0$ (top row), for which the equilibrium can be calculated analytically, and a regime with mutation rate $\mu=0.05$ (bottom row) which is explored by individual-based simulations. For the copying process, we assume that individuals $A$ and $B$ are chosen randomly. $A$ switches to $B$'s strategy with a probability given by $(1+\exp[s(P_A-P_B)])^{-1}$, where $s \geq 0$ corresponds to 'selection strength', see e.g. \cite{traulsen06}. 
If $All D$ competes with $WSLS$ the population is mostly in the defector's state, independent of population size and the mutation rate (A). However, once $E_\chi$ or $TFT$ is added, $WSLS$ succeeds for sufficiently large populations (B and C). Adding $All C$ only leads to minor changes in the stationary distribution (D and E). Parameters: $b=3$, $c=1$, $s=1$, and $\chi=2$.}
\end{figure}
\newpage

\begin{figure}[h!]
\centering
\includegraphics[height=5cm]{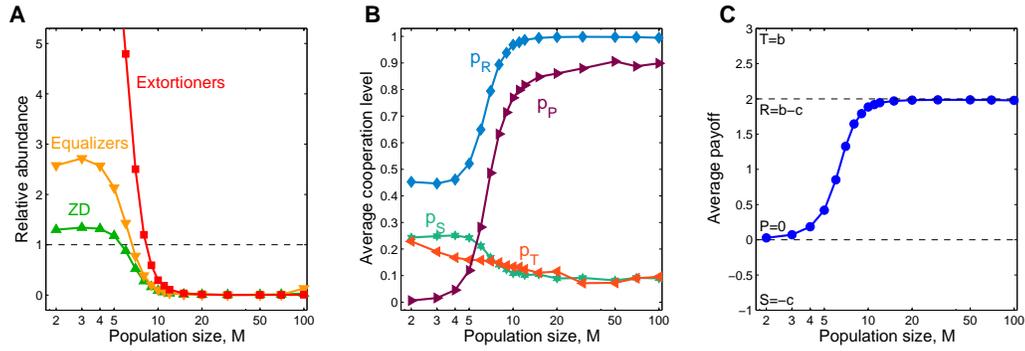}
\caption{Statistics of the evolutionary dynamics for memory-one strategies for a range of different population sizes. We have calculated (A) the relative abundance of extortioners, equalizers, and $ZD$ strategies, i.e. the time spent in a $\delta $-neighborhood, divided by the volume of that neighborhood; (B) the average strategy of the population; (C) the average payoff. Extortioners, equalizers and $ZD$-strategies are only favored for small population sizes. As the population size increases, individuals tend to apply $WSLS$-like strategies, and to cooperate only after mutual cooperation or mutual defection. As a result, the average payoff increases with population size. For the simulations, $10^7$ mutant strategies were randomly drawn from the space of memory-one strategies. The switch from a monomorphic population using strategy $X$ to a monomorphic population using strategy $Y$ occured with the probability of fixation of a single $Y$ mutant in a population of $X$-residents. Parameters: $b=3$, $c=1$, $\delta =0.1$ and $s=100$.}
\end{figure}

\begin{figure}[h!]
\centering
\includegraphics[height=6cm]{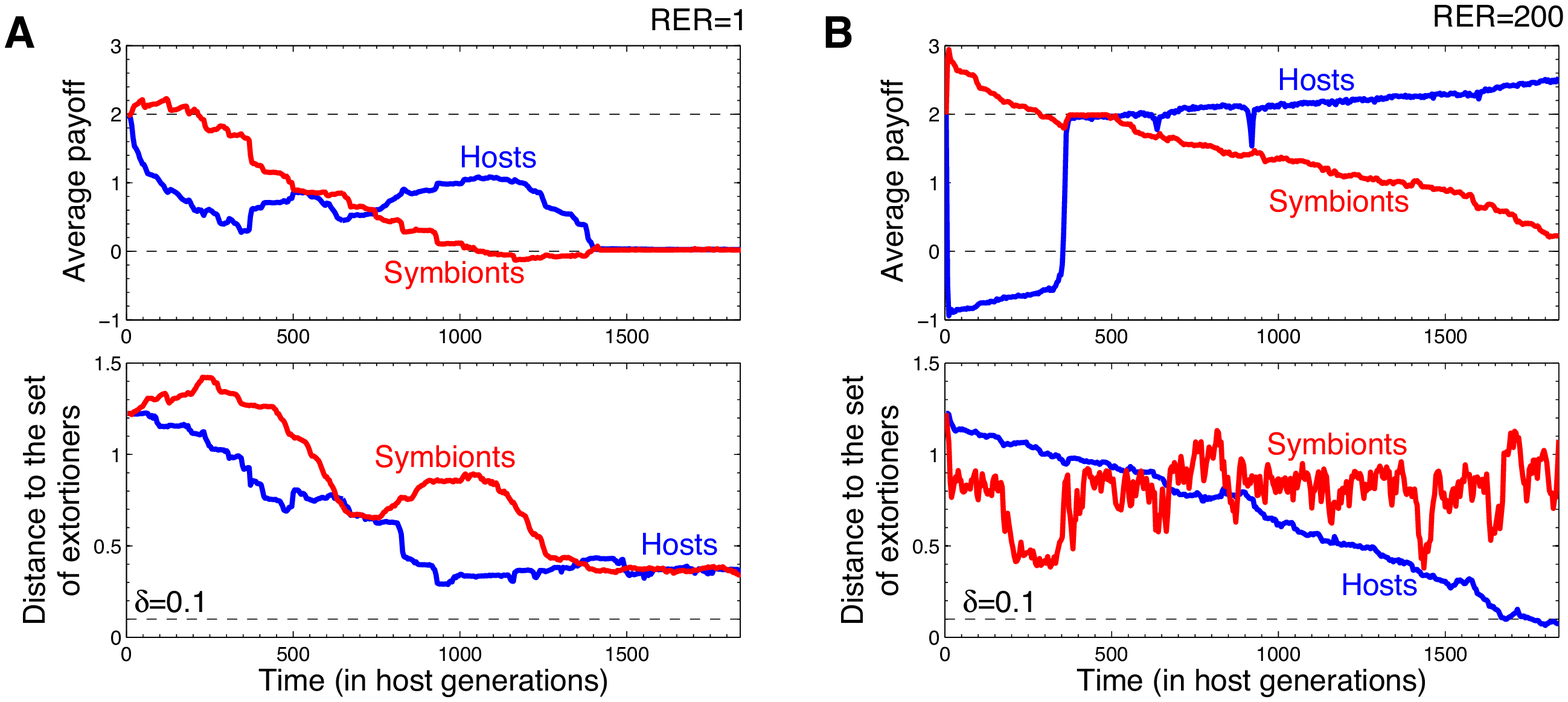}
\caption{Evolution of extortion in host-symbiont interactions. The graphs show two typical simulation runs for a population of $80$ hosts, each having a subpopulation of $25$ symbionts. For each simulation run, the upper graph shows the average payoff for each population, whereas the lower graph shows the Euclidean distance of each population to the set of extortioners (which can be at most $1.5275$). In the initial population all individuals cooperate unconditionally. The further evolution depends on the relative evolutionary rate: (A) If $RER=1$, both species converge towards $All D$, and no population is able to extort the other. (B) For $RER=500$, symbionts evolve much faster. In the short term, they can thus increase their average payoff by switching to a non-cooperative strategy. However, in the long term hosts apply extortion strategies to force their symbionts to cooperate. Eventually, the hosts' payoff exceeds $b-c$, whereas the symbionts' payoff is close to zero. To model the evolutionary process, we followed \cite{damore11}: Whenever a symbiont reproduces, its offspring remains associated with the same host. Whenever the host reproduces, the new host offspring acquires its symbionts from other hosts (horizontal transmission). Mutations occur with probability $\mu= 0.05$, by adding Gaussian noise to the memory-one strategy of the parent ($\sigma=0.05$). The process is run for $1,000$ host generations (corresponding to more than $10^6$ reproduction events for $RER=1$, and more than $10^9$ reproduction events for $RER=500$). The other parameters were $b=3$, $c=1$ and $s=10$.}
\end{figure}


\begin{thebibliography}{99}

\bibitem{rapoport65}
Rapoport A, Chammah A (1965) {\it The Prisoner's Dilemma}. (Ann Arbor, Univ Michigan Press).

\bibitem{trivers71}
Trivers R (1971) The evolution of reciprocal altruism. {\it Quarterly Rev  Biol} 46: 35-57.

\bibitem{aumann81}
Aumann R (1981) Survey of repeated games. In {\it Essays in Game Theory and Mathematical Economics in Honor of Oskar Morgenstern}, (Mannheim, Wissenschaftsverlag).

\bibitem{axelrod84}
Axelrod R (1984) {\it The Evolution of Cooperation} (Basic Books, New York). 

\bibitem{fudenberg86}
Fudenberg D, Maskin E (1986) The folk theorem in repeated games with discounting or with incomplete information. {\it Econometrica} 50: 533-554.

\bibitem{fudenberg90}
Fudenberg D, Maskin E (1990) Evolution and Cooperation in noisy repeated games. {\it American Economic Review} 80: 274-279.

\bibitem{nowak93}
Nowak MA, Sigmund K (1993) A strategy of win-stay, lose-shift that outperforms tit-for-tat in the Prisoner's Dilemma game. {\it Nature} 364: 56 - 58. 

\bibitem{kendall07}
Kendall G, Yao X, Chong, SY (eds) (2007) {\it The Iterated Prisoner's Dilemma: 20 Years On}. (Singapore: World Scientific).

\bibitem{trivers06}
Trivers R (2006) Reciprocal altruism: 30 years later. In {\it Cooperation in Primates and Humans: Mechanisms and Evolution}
(Kappeller PM, van Schaik CP eds.) (Berlin, Springer).

\bibitem{press12}
Press WH,  Dyson FD (2012) Iterated Prisoner's Dilemma contains strategies that dominate any evolutionary opponent. {\it Proc Nat Acad Sci USA} 109:

\bibitem{stewart12}
Stewart AJ,  Plotkin JB (2012) Extortion and Cooperation in the Prisoner's Dilemma. {\it Proc Nat Acad Sci USA} 109: 10134-10135.

\bibitem{boerlijst97}
Boerlijst MC, Nowak MA, Sigmund K (1997) Equal pay for all prisoners. {\it American Mathematical Monthly} 104: 303-307. 


\bibitem{bergstrom03}
Bergstrom CT, Lachman M  (2003) The Red King effect: when the slowest runner wins the race
{\it Proc Nat Acad Sci USA} 100: 593-598. 

\bibitem{sigmund10}
Sigmund K (2010) {\it The Calculus of Selfishness}. (Princeton Univ Press).

\bibitem{nowak90}
Nowak MA, Sigmund K (1990) The evolution of stochastic strategies in the Prisoner's Dilemma. {\it Acta Applicandae Math} 20: 247-265.


\bibitem{helbing92}
Helbing D (1992) A mathematical model for behavioural changes by pair interactions. Haag G, Mueller U, Troitzsch KG (eds), {\it Economic Evolution and Demographic Change. Formal Models in Social  Sciences} (Berlin, Springer) 330-348.

\bibitem{nowak06}
Nowak MA (2006) {\it Evolutionary Dynamics} (Cambridge MA, Harvard Univ Press).

\bibitem{nowak04}
Nowak MA, Sasaki A, Taylor C, Fudenberg D (2004) Emergence of cooperation and evolutionary stability in finite populations. {\it Nature} 428: 646-650.

\bibitem{imhof05}
Imhof L, Fudenberg D, Nowak MA (2005) Tit-for-tat or win-stay, lose-shift? {\it Journal of Theoretical Biology} 247: 574-580.

\bibitem{imhof10}
Imhof LA, Nowak MA (2010) Stochastic evolutionary dynamics of direct reciprocity. {\it Proc R Soc B} 277: 463-468.

\bibitem{traulsen06}
Traulsen A, Nowak MA, Pacheco J (2006) Stochastic dynamics of invasion and fixation. {\it Phys Rev E} 74 (1): 011909. http://dx.doi.org/10.1103/PhysRevE.74.011909

\bibitem{rand10}
Rand DG, Armao JJ, Nakamaru M, Ohtsuki H (2010) Anti-social punishment can prevent the co-evolution of punishment and cooperation. {\it J Theor Biol} 265: 624-632.

\bibitem{hilbe12}
Hilbe C,  Traulsen A (2012) Emergence of responsible sanctions without second-order free-riders, anti-social punishment or spite. {\it Scientific Reports} 2: 458.

\bibitem{nowak94}
Nowak MA, Sigmund K (1994) The alternating Prisoner's Dilemma. {\it J Theor Biol} 168: 219-226.

\bibitem{frean94}
Frean MR (1994)  The Prisoner's Dilemma without synchrony. {\it Proc Roy Soc London B} 257: 75-79.

\bibitem{damore11}
Damore JA, Gore J (2011) A slowly evolving host moves first in symbiotic interactions. {\it Evolution} 65:2391-2398.

\bibitem{doebeli98}
Doebeli M, Knowlton N (1998) The evolution of interspecific mutualisms. {\it Proc Nat Acad Sci USA} 95: 8676-8680. 

\bibitem{frean04}
Frean MR, Abraham ER (2004) Adaption and enslavement in endosymbiont-host associations. {\it Physical Review E} 69: 051913.

\bibitem{adami12}
Adami C, Hintze A (2012) Winning isn't everything: Evolutionary stability of Zero Determinant strategies. {\it preprint} http://arxiv.org/abs/1208.2666. 

\end{thebibliography}
\end{document}